\documentclass[useAMS,usenatbib,twocolumn]{mn2e}
\usepackage{amssymb}
\usepackage{graphicx}
\usepackage{lscape}
\usepackage{longtable}
\usepackage{array}
\newcommand\msun{$\rm\,M_\odot$}

\title[The IMF in giant elliptical galaxies: Top to Bottom]
{The (galaxy-wide) IMF in giant elliptical galaxies:\\ From top to bottom}

\author[C.~Weidner et al.]
{Carsten Weidner$^{1,2}$\thanks{E-mail: cweidner@iac.es}, Ignacio Ferreras$^{3}$, Alexandre Vazdekis$^{1,2}$, \newauthor Francesco La Barbera$^4$\\
$^{1}$Instituto de Astrof{\'i}sica de Canarias, Calle V{\'i}a L{\'a}ctea s/n, E38205, La Laguna, Tenerife,
Spain\\
$^{2}$Dept. Astrof{\'i}sica, Universidad de La Laguna (ULL), E-38206 La Laguna, Tenerife, Spain\\
$^{3}$Mullard Space Science Laboratory, University College London, Holmbury St Mary, Dorking, Surrey RH5 6NT\\
$^4$INAF--Osservatorio Astronomico di Capodimonte, I-80131 Napoli, Italy
}

\begin{document}
\bibliographystyle{aa}
\date{Accepted 2013 August 01. Received 2013 July 30; in original form 2013 June 12}

\pagerange{\pageref{firstpage}--\pageref{lastpage}} \pubyear{2013}

\maketitle

\label{firstpage}

\begin{abstract}
Recent evidence based independently on spectral line strengths and
dynamical modelling point towards a non-universal stellar Initial Mass
Function (IMF), probably implying an excess of low-mass stars in 
  elliptical galaxies with a high velocity dispersion. Here we show
that a time-independent bottom-heavy IMF is compatible neither with
the observed metal-rich populations found in giant ellipticals nor
with the number of stellar remnants observed within these systems. We
suggest a two-stage formation scenario involving a time-dependent IMF
to reconcile these observational constraints. In this model, an early
strong star-bursting stage with a top-heavy IMF is followed by a more
prolonged stage with a bottom-heavy IMF. Such model is physically
motivated by the fact that a sustained high star formation will bring
the interstellar medium to a state of pressure, temperature and
turbulence that can drastically alter the fragmentation of the gaseous
component into small clumps, promoting the formation of low-mass
stars. This toy model is in good agreement with the different
observational constrains on massive elliptical galaxies, such as age,
metallicity, $\alpha$-enhancement, M/L, or the mass fraction of the
stellar component in low-mass stars.
 \end{abstract}


%%%%%%%%%%%%%%%%%%%%%%%%%%%%%%%%%%%%%%%%%%%%%%%%%%%%%%%

\section{Introduction}
\label{se:intro}
One of the most fundamental properties of a stellar population is the
distribution with respect to mass at birth, i.e. the stellar initial
mass function (IMF). It is a highly important distribution function in
astrophysics, as stellar evolution is mostly determined by stellar
mass. The IMF therefore regulates the chemical enrichment history of
galaxies, as well as their mass-to-light ratios, and influences their
dynamical evolution. Studies of resolved stellar populations in the
Milky Way (MW) and the Magellanic clouds suggest that the IMF is
invariant over a large range of physical conditions like gas density
and metallicity \citep{Kr02,BCM10,KWP13}. This evidence motivated the
use of a fixed IMF for the description of the stellar populations of
whole galaxies, irrespective of their star formation history. However,
the underlying assumption that the IMF -- derived from and tested on
star cluster scales in the MW and nearby galaxies -- is the
appropriate stellar distribution function for more complex stellar
populations lies beyond present observational capabilities.

The advances of large-scale observational surveys comprising tens of
thousands of galaxies, as well as more detailed stellar modelling, led
to a number of results questioning the universality of the IMF
\citep[see, e.g.][]{HG:08,MWK09,LGT09,GHS10,CMA12,FBR13}.
Some of the results seem to be conflicting, especially in galaxies
involving high star formation rates. For example, \citet{GHS10} find
that the IMF becomes top-heavy in strong star-bursts while
\citet{FBR13} derive very bottom-heavy IMFs for massive elliptical
galaxies, which ought to have formed in star-bursts.  Such variations
in the IMF can have important implications in the derived star
formation histories and stellar masses of galaxies \citep{FVR13,FLB:13}.
In addition, chemical evolution models of lower mass spheroids,
  such as the bulges of the MW and M31 hint at a top-heavy IMF
  \citep{BKM07}. And some individual clusters like, e.g., M82-F, show
  possible top-heavy IMFs \citep{SG01} as well as massive globular clusters
  and ultra-compact dwarf galaxies \citep{DKB09,DKP12}. These pieces of 
  seemingly contradicting evidence could, instead, suggest that the evolution 
  of the IMF is much more complex, with a strong sensitivity to the local properties
  of the ISM, therefore correlating with mass, velocity dispersion and
  time.

%%%%%%%%%%%%%%%%%%%%%%%%%%%%%%%%%%%%%%%%%%%%%%%%%%%%%%%%%%%%%%%%%
\begin{figure}
\includegraphics[width=8cm]{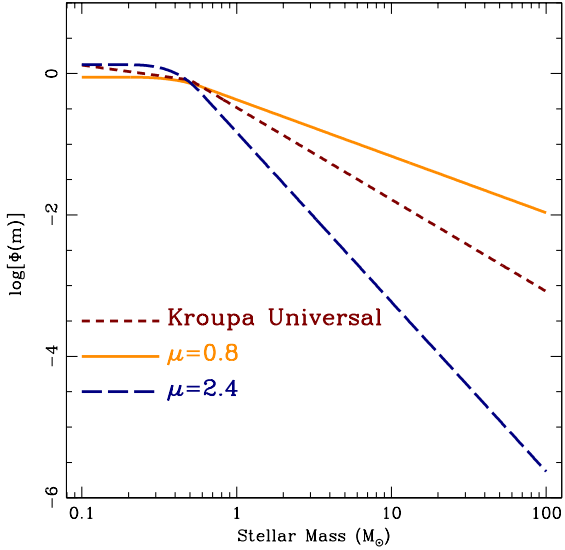}
\caption{Bimodal IMF functions used in this paper. We follow the
  prescription of \citet{VCP96}, which consists of a power law, with
  index $\mu$, for high masses, tapering off to low masses around
  0.4\msun. The choice of $\mu=1.3$ is almost indistinguishable from a
  \citet{Kr01} mass function.  }
\label{fig:IMF}
\end{figure}

In a volume-limited study of 260 early-type galaxies (hereafter ETGs)
via integral field spectroscopy and photometry, \citet{CMA12} find
that, regardless of the assumptions about the underlying dark matter
halos, the SDSS $r$-band mass-to-light ratios agree neither with the
assumption of a single slope Salpeter IMF \citep{Sal55}, nor with the
Kroupa-IMF \citep{Kr02}. \citet{CMA12} conclude that either a very
bottom-heavy (dominated by low-mass stars), or a very top-heavy IMF
(dominated by remnants) can explain the derived stellar M/L
ratios. \citet{FPV03} find a similar trend in the bulges of late-type
galaxies, and \citet{CVC03} suggest a bottom-heavy IMF as a viable
explanation to the low Calcium triplet abundances in massive
early-type galaxies. \citet{GK13} arrive at the same conclusion
 by studying the $g - z$ colours of seven massive elliptical galaxies
 and their metal-rich globular cluster systems.
\citet{VC10,VC12} provided strong evidence
towards a bottom-heavy IMF from the analysis of additional
gravity-sensitive features, such as Na8190 and FeH in massive
elliptical galaxies. Recently, \citet{L13} discussed the issue
  of a systematic underestimate of elemental abundances in the
  intracluster medium by a factor $>$2, and concluded that this
  mismatch could be reconciled by a non-standard IMF in massive
  early-type galaxies. The recent bone of contention has been
  presented by \citet{SL13}, where a single massive (strong lens)
  early-type galaxy does not appear to have a bottom-heavy IMF, based
  on a comparison between lensing mass and stellar mass from
  population synthesis models. However, as the authors suggest,
  compactness may be the driver of IMF variations
  \citep{Laesker:13,Conroy:13}. The lens galaxy explored by
  \citet{SL13} is significantly extended. Moreover, one should notice
  that gravi ty-sensitive line strengths do not provide tight
  constraints on the integrated M/L of a stellar population. In fact,
  as shown in \citet{FLB:13}, models that match equally well the
  observed line strengths (e.g.~single power-law and low-mass tapered
  IMF shapes) can have significantly different M/L \citep[see Fig.~4
    of][]{FBR13}. On the contrary, gravity-sensitive line strengths
  strongly constrain the fraction in low-mass stars at birth
  \citep[see Fig.~21 of][]{FLB:13}.

%%%%%%%%%%%%%%%%%%%%%%%%%%%%%%%%%%%%%%%%%%%%%%%%%%%%%%%
\begin{figure}
\includegraphics[width=85mm]{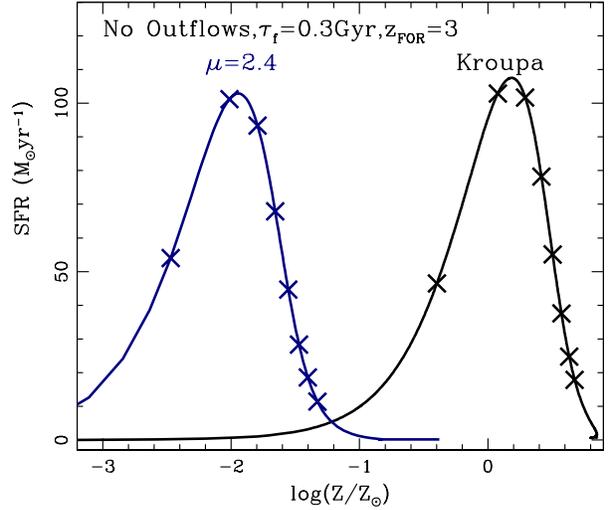}
   \caption{Metal enrichment history and star-formation history for
     three models assuming a fixed bimodal IMF, and a total stellar
     mass of 10$^{11}$\msun. In addition to a standard \citet{Kr01} IMF, we
     include the result for a bimodal IMF (see text for
     details) that corresponds to early-type galaxies with central
     velocity dispersion of $\sigma_0=200$\,km\,s$^{-1}$ ($\mu=2.4$),
     according to the relationship obtained in \citet{FLB:13} using a 
     hybrid method combining spectral fitting and targeted line
     strength analysis. The chemical evolution model has no outflows,
     with a gas infall time-scale of 0.3\,Gyr, and a formation redshift
     of z$_\mathrm{FOR}$=3. The crosses in both models, from left to
     right, mark ages from 0.25 to 2\,Gyr in steps of 0.25\,Gyr.}
\label{fig:metal}
\end{figure}

Although it is still early days -- both on the observational and
  modelling sides -- for a robust confirmation of a bottom-heavy IMF
  in massive/high velocity dispersion/compact early-type galaxies, the
  evidence is strong enough to ponder about the consequences of such a
  scenario on the underlying populations and chemical enrichment.
This paper is structured as follows. In \S\ref{se:cons} the
implications of a bottom-heavy IMF on the chemical evolution and the
amount of remnants in massive elliptical galaxies is discussed, while
in \S\ref{se:model} the chemical evolution model used is presented,
along with the results for three choices, using different
IMFs. Finally, in \S\ref{se:diss} the results of this work are
discussed.\\

%%%%%%%%%%%%%%%%%%%%%%%%%%%%%%%%%%%%%%%%%%%%%%%%%%%%%%
\section{Issues from a time-independent bottom-heavy IMF}
\label{se:cons}

%%%%%%%%%%%%%%%%%%%%%%%%%%%%%%%%%%%%%%%%%%%%%%%%%%%%%%%%%%%%%%%%%
\begin{figure}
\includegraphics[width=85mm]{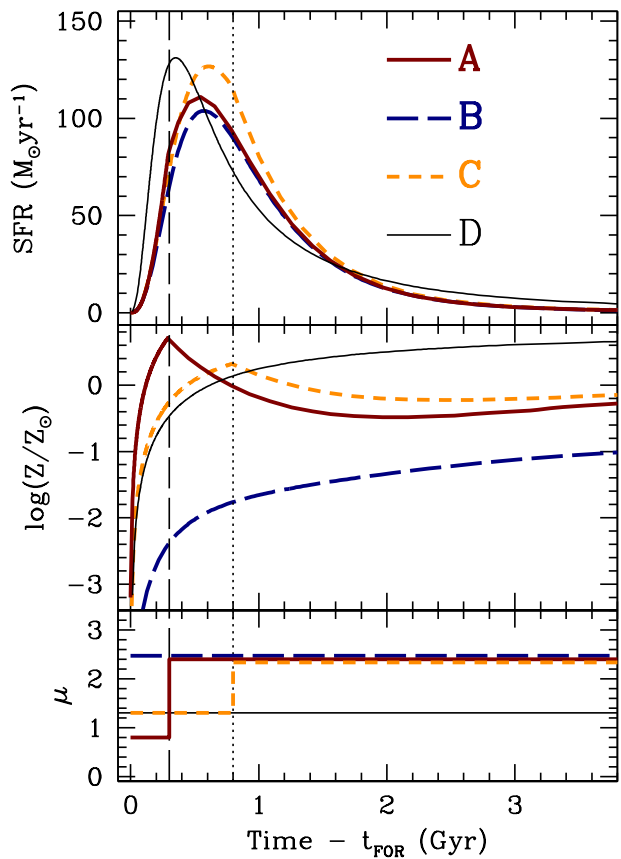}
\caption{Comparison of four models of chemical enrichment, as
  described in Tab.~\ref{tab:models}. All models are chosen to
  reproduce the age distribution of a massive early-type galaxy
  (i.e. early formation and narrow distribution of ages to
  create the old, $\alpha$-enhanced populations observed in 
  these galaxies). The top panel shows the star formation rate,
  assuming a total stellar mass of 10$^{11}$\msun. The middle panel
  tracks the gas-phase metallicity. The bottom panel illustrates
  the evolution of the slope of the IMF. A bimodal IMF is
  assumed in models A, B, C, whereas model D corresponds to a 
  time-invariant Kroupa IMF (see text for details).}
\label{fig:toy}
\end{figure}

\subsection{Chemical enrichment}

One important problem not sufficiently discussed in the literature is
that a bottom-heavy IMF, especially as steep as proposed by
\citet{FBR13}, leads to a severe under-production of metals in massive
galaxies. These systems are known to have solar or even super-solar
metallicities \citep[e.g.][]{TFW00,ABR07}, but an IMF with a power-law
index\footnote{In this paper we use the bimodal IMF proposed by
  \citet{VCP96} and in Appendix A of \citet{VCG03}, which consists of a power law, with index $\mu$, for
  high masses, tapering off to low masses around 0.4\msun\ (see
  Fig.~\ref{fig:IMF}). With this function, a $\mu=1.3$ case is
  indistinguishable from a \citet{Kr01} or \citet{Chab:03} IMF
  \citep[see, e.g., Fig.~3 of][]{FLB:13}.} of $\mu$ = 2.4 for stars
above 1\msun\,only results into $10^8$ stars more massive than
8\msun\,for a galaxy of 10$^{11}$\msun\,while a \cite{Kr01} IMF gives
$10^9$ such stars -- a factor of 10 times higher. This reduction of the
number of massive stars carries severe consequences for the chemical
enrichment of the galaxy. For instance, the amount of O$^{16}$
released by massive stars for the $\mu$ = 2.4 case, and extrapolating
the \citet{NTU06} yields up to stellar masses of 100\msun\,, results
in a 7 times lower than solar oxygen abundance. This is a simple,
back-of-the-envelope estimate that can be significantly affected by a
process of continuous gas infall and enrichment. In
Sec.~\ref{se:model} we discuss in more detail the issue of galactic
chemical enrichment by exploring a set of toy models, confirming the
inability of a time-independent, bottom-heavy IMF to explain the
metal-rich populations of massive ETGs.

%%%%%%%%%%%%%%%%%%%%%%%%%%%%%%%%%%%%%%%%%%%%%%%%%%%%%%
\begin{figure}
\includegraphics[width=8cm]{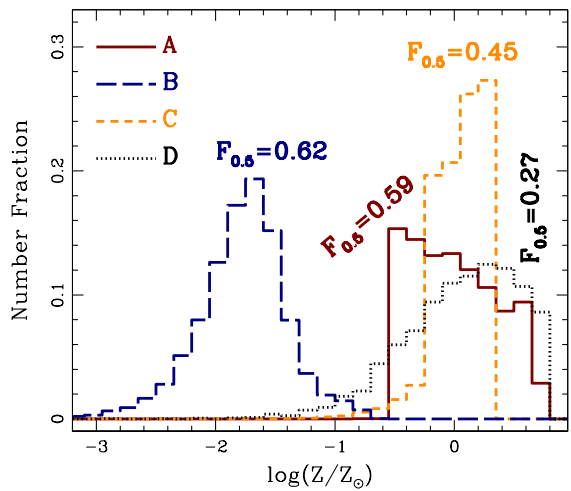}
\caption{Stellar metallicity distribution corresponding to the 
  models shown in Fig.~\ref{fig:toy}. The number for each example
  corresponds to the mass fraction in stars below 0.5\,M$_\odot$.}
\label{fig:toy2}
\end{figure}

%%%%%%%%%%%%%%%%%%%%%%%%%%%%%%%%%%%%%%%%%%%%%%%%%%%%%%%%%%%%%%%
\subsection{Stellar remnants}

An independent test of the shape of the IMF of massive early-type
galaxies comes from the stellar remnants of the population. The
assumption of a time-independent, bottom-heavy IMF naturally leads to
a low number of low-mass X-ray binaries (LMXB). As these systems
consist of a neutron star and a low-mass companion, their number is
directly influenced by the lower number of remnants expected from such
an IMF. Other parameters impact the number of LMXBs especially in
  globular clusters (GC) as well. Metallicity in special
  \citep{F06,MKD12,KFI13}, and also the initial density of the cluster
  forming cloud \citep{MKD12} can change the number of LMXBs formed in
  GCs. While the number of LMXBs in metal-poor GCs is about three
  times lower than in metal-rich GCs \citep{KFI13} the fraction of
  metal-poor GCs with LMXBs is only about two times lower than for
  metal-rich GCs \citep{KFB09}. As almost all galaxies host both types
  of GCs, this effect his further diluted in respect to the much larger
  impact expected of a very bottom-heavy IMF.

 Several studies of LMXBs in elliptical galaxies show that these
 exotic binaries are not underrepresented in elliptical galaxies
 \citep{KMZ02,F06,KFB09,KFI13}.  Table~\ref{tab:lmxbs} summarizes the
 results from the literature for the LMXBs found in the GCs of several
 nearby galaxies of different types. The fraction of GCs with LMXBs is
 found to vary in ETGs between 3\% and $\sim$20\% when correcting to
 the same x-ray detection limit. For the three spiral galaxies in the
 sample, the Milky Way, M31 and M81, $\sim$5--10\% of GCs have
 LMXBs. Taking into account the uncertainties of the stellar mass
 determination, these numbers compare well with the hypothesis that
 all these galaxies have similar IMFs. Additionally, very little
 impact seem to come form the total luminosity (and hence mass) or 
 morphological type, a conclusion shared by \citet{ZGV11}.  For
 field LMXBs, less information is available. When roughly correcting to
 the same x-ray luminosity completeness limit of $L_\mathrm{x-ray,
   lim}$ $\ge$ 10$^{36}$ erg s$^{-1}$, all but one galaxy (Maffei 1)
 have an amount of field LMXBs of about 30 to 120, similar to the 100
 field LMXBs in the MW.  These numbers imply that massive
early-type galaxies cannot have a fixed bottom-heavy IMF, as this
would lead to much fewer LMXBs than observed (see
\S~\ref{se:diss}). However, a more quantitative study would
  require a larger sample of LMXBs in the field and globular cluster
  populations of different types of galaxies, also
  including dwarf galaxies, and would need to focus on the field
  LMXB population. This would also require detailed population
  modelling of the galaxies to separate the old and young contribution
  in spirals \footnote{High-mass x-ray binaries (HMXBs) have not
    been included here as they are only found in systems with young
    and massive stars which are scarce in ETGs. As they probe very
    different parts of the IMF with respect to the stellar companion
    the observational results on HMXBs are not readily comparable to
    the LMXB results.}

%%%%%%%%%%%%%%%%%%%%%%%%%%%%%%%%%%%%%%
%%%%%%%%%%   TABLE 1   %%%%%%%%%%%%%%%
%%%%%%%%%%%%%%%%%%%%%%%%%%%%%%%%%%%%%%
\begin{table*}
\begin{center}
\begin{tabular}{cccccccccc}
\hline
Name&type&$L_\mathrm{V}$&fraction of GCs&field LMXBs&$L_\mathrm{x-ray, lim}$&Ref.\\
&&10$^{10}$ $L_\odot$&with LMXBs&&10$^{36}$ erg s$^{-1}$&\\
\hline
NGC 1399&E1pec/cD&8.1&6&-&1&(1)\\
NGC 1399&E1pec/cD&8.1&4&-&10&(2)\\
NGC 1404&E1&3.8&3&-&1&(1)\\
NGC 1427&E5&1.1&6&-&1&(1)\\
NGC 3379&E1&1.9&12&26&6&(3)\\
NGC 3379&E1&1.9&12-20&46&4&(4)\\
NGC 4278&E1-2&2.2&14&42&15&(3)\\
NGC 4278&E1-2&2.2&10-17&74&6&(4)\\
NGC 4365&E3&9.3&8&-&1&(1)\\
NGC 4374&E1&9.3&3&-&1&(1)\\
NGC 4472&E2&17.6&9&-&1&(1)\\
NGC 4472&E2&17.6&4&-&10&(2)\\
NGC 4621&E5&4.6&5&-&1&(1)\\
NGC 4649&E2&9.9&9&-&1&(1)\\
NGC 4697&E6&1.7&7&44&15&(3)\\
NGC 4697&E6&1.7&8-10&61&5&(4)\\
Cen A&S0pec&2.2&7-17&120&0.6&(4)\\
Maffei 1&S0pec&0.02&20&33&3&(4)\\
M 31&SAb&3.3&2-3&-&10&(2)\\
M 31&SAb&3.3&21&166&0.04&(4)\\
M 81&SAab&2.0&8-12&133&0.7&(4)\\
MW&SBc&$\approx$1.0&8&$\approx$100&1&(5)\\
MW&SBc&$\approx$1.0&1-4&-&10&(2)\\
\hline
\end{tabular}\\
\end{center}
(1) \citet{KFI13}, (2) \citet{KMZ02}, (3) \citet{KFB09}, (4) \citet{ZGV11}, (5) \citet{GGS02,H96}
\caption{Properties of LMXB population in globular clusters of various galaxies. The 'type' classification and the $L_\mathrm{V}$ values are from NASA/IPAC extragalactic database (http://ned.ipac.caltech.edu). In the cases with a range for the fraction of GCs with LMXBs completeness corrections for GCs and LMXBs have been applied.}
\label{tab:lmxbs}
\end{table*}
%%%%%%%%%%%%%%%%%%%%%%%%%%%%%%%%%%%%%%%%%%%%%%%%%%%%%%%%%%%

%%%%%%%%%%%%%%%%%%%%%%%%%%%%%%%%%%%%%%
%%%%%%%%%%   TABLE 2   %%%%%%%%%%%%%%%
%%%%%%%%%%%%%%%%%%%%%%%%%%%%%%%%%%%%%%
\begin{table*}
\begin{center}
\begin{tabular}{cccccccccccc}
\hline
Model & $\tau_f$ & B$_{\rm out}$ & $\mu_1$ & $\mu_2$ & $\Delta t_{\rm IMF}$ &
$\langle$Age$\rangle$  & $\langle\log Z/Z_\odot\rangle$ & $[\alpha/{\rm Fe}]^1$ & F($<$0.5M$_\odot$) & M/L & \# Remnants\\
 & Gyr & & & & Gyr & Gyr & & & & SDSS $r$ & per star\\
\hline
A & 0.3 & 0.0 & $0.8$ & $2.4$ & $0.3$ & $10.29$ & $+0.01$ & $+0.26$ & $0.59$ & $4.93$ & 1/139\\
B & 0.3 & 0.0 & $2.4$ & $2.4$ & --- & $10.26$ & $-1.78$ & $+0.26$ & $0.62$ & $2.21$ & 1/2419\\
C & 0.3 & 0.0 & $1.3$ & $2.4$ & $0.8$ & $10.27$ & $+0.03$ & $+0.26$ & $0.45$ & $4.33$ & 1/367\\
D & 0.1 & 0.3 & $1.3$ & $1.3$ & --- & $9.94$ & $+0.06$ & $+0.25$ & $0.27$ & $3.18$ & 1/145\\
\hline 
\end{tabular}\\
\end{center}
$^1$ We follow the proxy between the time to form half of the stellar mass
and [$\alpha$/Fe] from \citet{IGDR:11}.
\caption{Properties of the three toy chemical enrichment models shown
  in Figs.~\ref{fig:toy} and \ref{fig:toy2}. All models begin at
  z$_{\rm FOR}=3$, with high star formation efficiency (C$_{\rm
    eff}=20$). The parameter $\Delta t_{\rm IMF}$ is the time lapse
  when the system has an IMF slope $\mu_1$, changing abruptly to
  $\mu_2$ afterwards. The average values of age and metallicity are
  weighted by the star formation rate.}
\label{tab:models}
\end{table*}
%%%%%%%%%%%%%%%%%%%%%%%%%%%%%%%%%%%%%%%%%%%%%%%%%%%%%%%%%%%

%%%%%%%%%%%%%%%%%%%%%%%%%%%%%%%%%%%%%%%%%%%%%%%%%%%%%%%
\section{Two-stage formation scenario}
\label{se:model}

To address these problems, we make use of a simple modellisation of
chemical enrichment as set out in \citet{FS:00a,FS:00b}. In essence,
the buildup of the stellar component of a galaxy is described by four
parameters: a gas infall timescale ($\tau_f$), a star formation
efficiency ($C_{\rm eff}$), that follows a Schmidt law, a formation
redshift ($z_{\rm FOR}$) at which the whole process starts, and a
fraction of gas ejected in outflows ($B_{\rm out}$). We refer the
reader to those references for details. We note that these
models feed star formation through the infall of primordial
(i.e. zero metallicity) gas. Therefore, no pre-enrichment is
assumed. In \citet{FS:00a,FS:00b} it was shown that the model
could reproduce the ages and metallicities of early-type galaxies,
with highly efficient star formation and short-lived infall
for the most massive galaxies.

For our purposes, we choose
the typical values required to explain the old, metal-rich populations
found in massive early-type galaxies: a short infall timescale
($\tau_f=0.3$\,Gyr), an early start ($z_{\rm FOR}=3$), a high
efficiency ($C_{\rm eff}=20$), and negligible outflows ($B_{\rm
  out}=0$). Fig.~\ref{fig:metal} shows the relationship between the
star formation rate and the gas-phase metallicity -- giving a direct
estimate of the stellar metallicity distribution -- assuming a
time-independent IMF. Notice that at the high values of $\mu$ needed
by the line strength analysis of the SDSS massive early-type galaxy
data \citep[$\mu\gtrsim 2$,][]{FBR13,FLB:13}, it is not possible to
obtain enough metals to reach the solar/super-solar metallicities
found in these galaxies, confirming the simple estimate made in the
previous section. Therefore, any model assuming a time-independent IMF
is not capable of explaining {\sl both} the metal-rich content and the
excess fraction in low-mass stars.

%%%%%%%%%%%%%%%%%%%%%%%%%%%%%%%%%%%%%%%%%%%%%%%%%%%%%%%%
\begin{figure*}
\includegraphics[width=14cm]{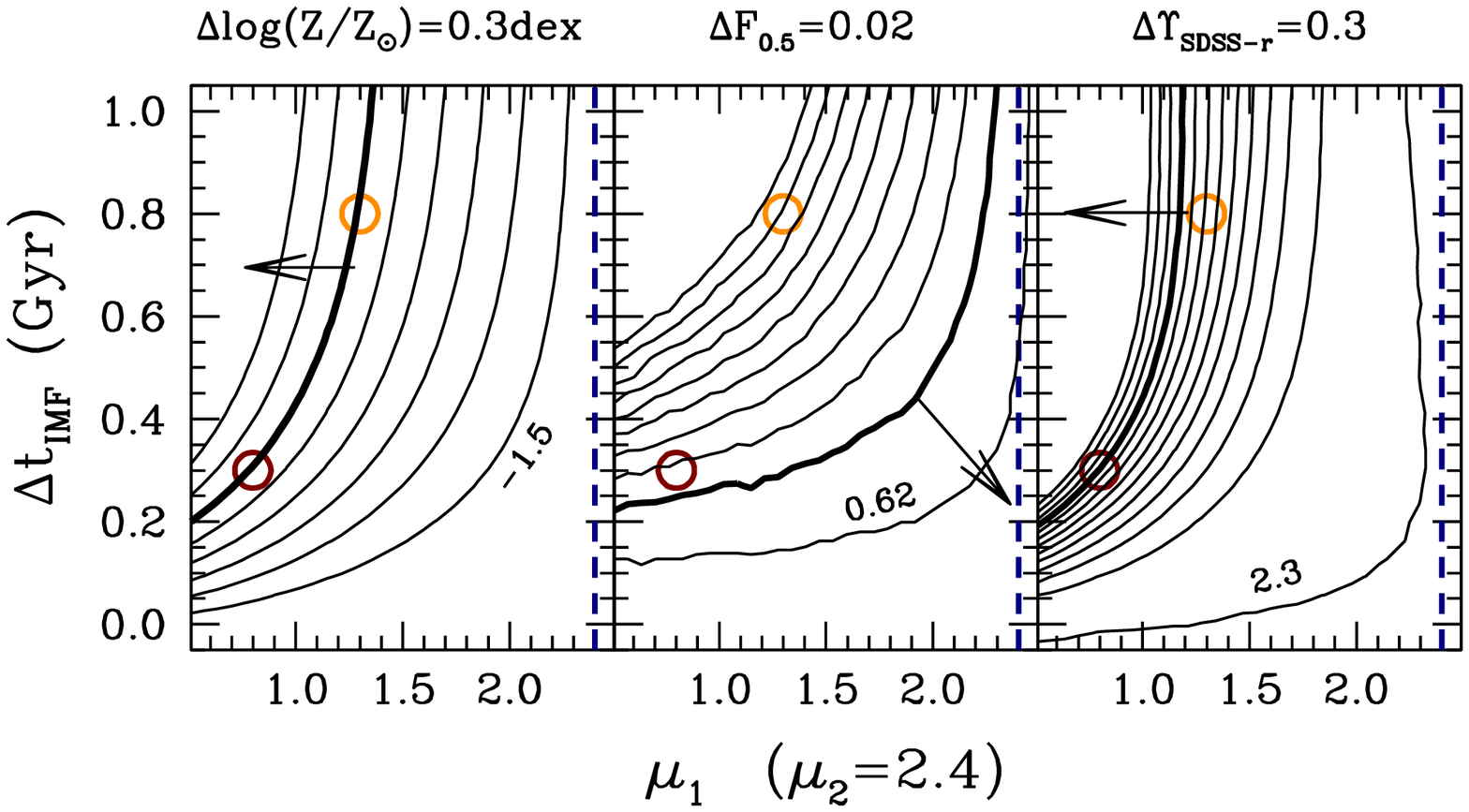}
\caption{Contours of average stellar metallicity (left); fraction
  in low-mass stars (middle) and SDSS-r M/L (right) for a number of
  models exploring the initial slope of the bimodal IMF (horizontal
  axis) and the switching time between IMFs (vertical line). The slope
  of the IMF during the second stage is fixed at 2.4 for all models.
  The thick contour corresponds to solar metallicity (left);
  F$_{0.5}$=0.6 (middle) and $\Upsilon_r=5$ (right). The arrows in
  each panel give the direction of increase, with the step per contour
  level given in the top part of each panel. The lower red (upper
  orange) circle marks the position of model A (C). Model B is
  represented by the vertical blue dashed line
  (i.e. $\mu_1=\mu_2=2.4$).}
\label{fig:conts}
\end{figure*}

To circumvent this problem, we follow the approach of
\citet{VCP96,VPB97}, namely invoking a time dependent IMF. We assume
that the formation of a massive galaxy starts with a very efficient
star formation, quickly achieving high star formation rates,
triggering a top-heavy IMF as expected of starbursting systems
\citep{WKPA:11,KWP13}.  However, after a period of time, one would expect
that the physical conditions of the gas are such that the IMF turns to
a bottom-heavy shape. In our toy model, we assume a $\mu=0.8$ initial
slope, followed by a $\mu=2.4$ distribution after 0.3\,Gyr 
(see Fig.~\ref{fig:IMF}).
Figs.~\ref{fig:toy} and \ref{fig:toy2} show that this simple model
(Model A, red solid lines) is capable of explaining the age and metallicity
distribution of a massive ETG, along with a high fraction in low-mass
stars. For reference, we also include the bottom-heavy,
time-independent case (Model B, blue long-dashed lines), and a time-dependent
case, where the original distribution follows a Kroupa/Chabrier
function (Model C, orange short-dashed lines), with the transition to a
bottom-heavy IMF occurring at a later time (to allow for the buildup
of metallicity to solar levels). Note that all three cases give a very
similar star formation history (Fig.~\ref{fig:toy}, top panel),
producing the expected old, $\alpha$-enhanced populations found in
massive ETGs \citep[see, e.g.][]{TFW00,Thomas:05,IGDR:11}.
An additional model D (black lines) is included, corresponding
to a fixed Kroupa-like IMF ($\mu$=1.3). In this case, to avoid a
high super-solar metallicity, we allowed for some outflows and shortened
the gas infall timescale, to obtain a model with similar age and metallicity
values as in the previous ones.
Tab.~\ref{tab:models} gives additional information about these models.
The $[\alpha/{\rm Fe}]$ values, which are indicators for the length of
the star-formation period \citep{TFW00}, are derived following the
relationship between $\alpha$-enhancement and T$_{M/2}$, defined as
the time lapse to form one half of the final stellar mass of the
galaxy \citep[see Eq.~2 in][]{IGDR:11}. We note that these estimates
of $[\alpha/{\rm Fe}]$ give a rough approximation, since the original
calibration is 
based on a Kroupa IMF, the age distributions are determined via
spectral fitting, and the relationship present a scatter of
$\sim$0.1\,dex.  It should also be noted here that the first
population only constitutes $\lesssim$ 10\% of the present-day light of the
galaxy \citep[see fig.~21 in][]{VCP96} and therefore its weight to the
overall galaxy spectral energy distribution is expected to be
negligible. To summarize, all models give average ages and
[$\alpha$/Fe] compatible with the observations, but the
fixed bottom-heavy IMF (model B), yield too low metallicities, and the
fixed Kroupa-like IMF (Model D), produce too few low-mass stars.

Finally, the last column in Tab.~\ref{tab:models} shows the expected
number of remnants (per star in the galaxy), a quantity directly
related to the population of LMXBs. We note that a \cite{Kr01} IMF
produces one remnant per 145 stars, a result that is similar to our
Model A, whereas a time-independent, bottom-heavy IMF (model B) 
gives a much lower number of remnants, in conflict with the observations
of LMXBs in massive ETGs.

Hence, models A and C are equally acceptable, as they produce
  results compatible with the observations. A degeneracy is therefore
  expected between the IMF initial slope ($\mu_1$) and the
  ``switching'' time $\Delta t_{\rm IMF}$.  Fig.~\ref{fig:conts}
  explores the degeneracy in more detail, by showing contours in this
  2D parameter space for the stellar metallicity (left); fraction in
  low-mass stars (middle) and stellar M/L in the SDSS-$r$ band
  (right). Note that for all models, the IMF slope at $t>\Delta t_{\rm
    IMF}$ is fixed at $\mu_2=2.4$ for simplicity. For all models
  explored, the fraction in low-mass stars is rather homogeneous,
  around $F_{0.5}\sim 0.6$. There is a plateau at low M/L in the
  bottom-right corner of the 2D parameter space. It is caused by a
  combination of a too bottom-heavy initial IMF slope, and the
  expected low metallicities from such models. The region of interest
  regarding the observations is located in the upper-left corner of
  these panels, i.e. a top-heavy initial slope, $\mu_1$, followed by a
  bottom-heavy IMF after $\Delta t_{\rm IMF}\gtrsim 0.3$\,Gyr.

%%%%%%%%%%%%%%%%%%%%%%%%%%%%%%%%%%%%%%%%%%%%%%%%%%%%%%%
\section{Discussion \& Conclusions}
\label{se:diss}

The different pieces of observational evidence about the IMF in
massive ETGs seem to be conflicting at a first glance. The combination
of line-strengths and mass-to-light ratios imply a bottom-heavy
IMF \citep{FLB:13,CMA12}. However, the solar or super-solar metallicity of these systems
does require at least a Kroupa-like IMF, a result supported as well by
the observed number of low-mass X-ray binaries. Although a
time-independent, bottom-heavy IMF is clearly in tension with the
metal-rich populations, it is necessary to account for the high
fraction in low-mass stars observed through the presence of the
gravity-sensitive line strengths \citep{CVC03,VC10,FBR13}. In
\citet{FLB:13}, it is estimated that over 50\% of the stellar mass
created in a massive ETG should be in the form of $\lesssim$0.5\msun\ stars. 
A possible solution, presented in this
paper, involves a two-stage galaxy-formation model\footnote{To avoid
  confusion, we emphasise that these two stages relate to the early
  formation of the core of a massive galaxy, and have nothing to do
  with the two phases of formation of, e.g. \citet{Oser:10}, where
  reference is made to the formation of the core and the outer region
  of a massive galaxy through cosmic history.} with a variation of the IMF. The case of a
non-universal IMF has already been considered since the first studies
of star formation in galaxies \citep[e.g.][]{S63}, and a number of
later works have invoked changes in the IMF to explain the properties
of elliptical galaxies \citep[see,
  e.g.,][]{Worthey:92,Elbaz:95,VCP96}.  This simple two-stage approach
serves to illustrate how a time-dependent IMF can explain the
different observational data. A more realistic model should assume a
smoother transition. 
%However, we note that \citet{OTS05} find a very
%sharp change in the fragmentation behaviour of molecular clouds when
%considering metallicities above and below [Z/H]=$-5$ which would
%indicate a rapid change of the power-law index.

In our model, during the first stage, a small fraction ($\sim$10\% of
the population by mass) form in a very short burst, lasting
$\lesssim$0.3\,Gyr, with a top-heavy IMF (Fig.~\ref{fig:toy}, red
lines). This first stage is very efficient at building up a metal-rich
ISM, and also results in a high injection of energy from type II
supernov\ae.  This process is followed by a second stage where the
bulk of the stellar mass is formed in about 1\,Gyr, with a
bottom-heavy IMF. Although no theory of star formation has been
capable so far of explaining the most fundamental properties of the
IMF from the physics of the ISM, we can motivate this model, as after
a sustained period with a high star formation rate, one would expect
that a highly turbulent ISM, with a very high velocity dispersion
would be conducive to enhanced fragmentation
\citep{HChab:09,Hopkins:13}. Such a scenario is also consistent
with the results of \citet{RVC12}, where a comparison between
broadband photometry of SDSS galaxies and a wide Monte Carlo library
of synthetic models found very few compatible solutions, often
involving a superposition of two old populations with a top-heavy and
a bottom-heavy IMF, i.e. similar to our case A. Fig.~\ref{fig:toy2}
shows that such a model (red solid line) creates a distribution of
metallicities compatible with the observations, and without the
low-metallicity tail in the distribution, typical of closed box models
with a time-independent IMF. The average metallicities quoted in
Tab.~\ref{tab:models} for models A and C are compatible with those
obtained in the line strength analysis of the most massive ETGs in
\citet{FLB:13}. The models A, B and C all roughly agree on the
fraction by mass of the population in stars with mass below 0.5\msun,
quoted in Fig.~\ref{fig:toy2} as F$_{0.5}$. A comprehensive analysis
of line strengths and spectral fitting \citep{FLB:13} shows that an
elliptical galaxy with $\sigma_0=200$\,km\,s$^{-1}$ -- which is the
fiducial case considered here -- requires a fraction
$\gtrsim$50\%. Note that models A and C give very similar
observational constraints, reflecting an inherent degeneracy between
the IMF slope difference of the two phases ($\mu_1-\mu_2$), and the
transition timescale ($\Delta t_{\rm IMF}$).  Notwithstanding this
degeneracy (see Fig.~\ref{fig:conts}), the purpose of this paper
is to prove that the case $\mu_1=\mu_2$ (i.e. models B and D) is
readily ruled out by the observations. As can be seen in
  Tab.~\ref{tab:models}, model B results in a far too low mean
  metallicity and too few remnants in order to explain LMXBs, whereas
  a Kroupa-like, time-independent IMF is not capable of accounting
  for the high presence of low-mass stars.
  In addition, the low
fraction of stellar mass created during the top-heavy stage implies
that the M/L of our model (which uses a bimodal IMF) is fully
compatible with dynamical mass estimates \citep[see Fig.~21
  of][]{FLB:13}. It is worth emphasising  that a model with {\sl
  only} a high star formation efficiency is not capable of explaining
all the observational constraints in massive ETGs. A variation of the
IMF is a further requirement, as shown here.

Note that a theory based on empirical relations exists which allows
variations of the galaxy-wide IMF with the star-formation rate of a
galaxy -- the integrated galactic IMF 
\citep[IGIMF,][]{KW03,WK05a,KWP13}. In this paper we extend the original
concept of the IGIMF -- which uses the star formation rate as a {\sl
  proxy} of the physical conditions that lead to a change in the IMF
-- to include the cumulative effect that a previous stage sustaining a 
high SFR can have on the IMF. 

We presented strong evidence against the hypothesis of a time-
independent, bottom-heavy IMF in massive galaxies. An alternative
scenario is proposed, where the observational data can be explained by
a two-stage formation process, involving a variation of the IMF during
the star formation history of a massive galaxy. Although a complete
theory of star formation is missing at present, such a scenario can be
motivated by the fact that a sustained high SFR will undoubtedly alter
the physical conditions of the interstellar medium, leading to
significant differences in the fragmentation process of gas clumps
into cores and then stars. This paper is not meant to fully solve the
problem but, rather, to give a plausible interpretation that detailed
theoretical models of star formation should address.

%%%%%%%%%%%%%%%%%%%%%%%%%%%%%%%%%%%%%%%%%%%%%%%%%%%%%%%

\section*{Acknowledgements}
We thank Ignacio Trujillo, Russell Smith, Pavel Kroupa, R{\'e}mon Cornelisse and
Fabien Gris{\'e} for interesting discussions. The useful comments and
suggestions from the referee are duly acknowledged.  This research has made use of the NASA/IPAC Extragalactic Database (NED) which is operated by the Jet Propulsion Laboratory, California Institute of Technology, under contract with the National Aeronautics and Space Administration. Furthermore, this work 
has been supported by the Programa Nacional de Astronom{\'i}a y Astrof{\'i}sica
of the Spanish Ministry of Science and Innovation under grant
AYA2010-21322-C03-02.

%%%%%%%%%%%%%%%%%%%%%%%%%%%%%%%%%%%%%%%%%%%%%%%%%%%%%%%
\bibliography{mybiblio}

\bsp
\label{lastpage}
\end{document}